\documentclass{article}
\usepackage{amsmath}
\usepackage{amsfonts}
\usepackage{amssymb}

\begin{document}

\begin{center}
{\large \textsc{Quantum effects near future singularities}} \vskip8mm \textbf{John D. Barrow}$^{\,a,}$\footnote{%
E-mail: jdb34@damtp.cam.ac.uk}, \quad \textbf{Ant\^{o}nio B. Batista}$%
^{\,b,} $\footnote{%
E-mail: abrasilb918@gmail}, \quad \textbf{J\'{u}lio C. Fabris}$^{\,b,}$%
\footnote{%
E-mail: fabris@pq.cnpq.br}, \\[0pt]
\textbf{Mahouton J.S. Houndjo}$^{\,c,}$\footnote{%
E-mail: sthoundjo@yahoo.fr} \quad \textbf{and} \quad \textbf{Giuseppe Dito}$%
^{\,d,e,}$\footnote{%
E-mail: giuseppe.dito@u-bourgogne.fr} $\,,\,\,$ \vskip8mm $^{\,a}$ DAMTP,
Centre for Mathematical Sciences, University of Cambridge, UK\\[0pt]
$^{\,b}$ Departamento de F\'{\i}sica, Universidade Federal do Esp\'{\i}rito
Santo, ES, Brazil \\[0pt]
$^{\,c}$ Instituto de F\'{\i}sica, Universidade Federal da Bahia, Ba, Brazil 
\\[0pt]
$^{\,d}$ Instituto de Matem\'{a}tica, Universidade Federal da Bahia, Ba,
Brazil \\[0pt]
$^{\,e}$ Institut de Math\'ematiques de Bourgogne, Universit\'e de
Bourgogne, Dijon, France\\[0pt]
\end{center}

\vskip 12mm


\begin{quotation}
\noindent
General relativity allows a variety of future singularities to occur in the
evolution of the universe. At these future singularities, the universe will
end in a singular state after a finite proper time and geometrical
invariants of the space time will diverge. One question that naturally
arises with respect to these cosmological scenarios is the following: can
quantum effects lead to the avoidance of these future singularities? We
analyze this problem considering massless and conformally coupled scalar
fields in an isotropic and homogeneous background leading to future
singularities. It is shown that near strong, big rip-type singularities,
with violation of the energy conditions, the quantum effects are very
important, while near some  milder classes of singularity like the sudden
singularity, which preserve the energy conditions, quantum effects are
irrelevant. \newline
\end{quotation}

\vskip 4mm

\section{Introduction}

It is general believed that today the Universe is in a stage of accelerated
expansion. The primary evidence for this accelerated phase of cosmic
evolution came from the use of the supernova type Ia as standard candles to
measure distances in the universe. Supernova are very bright objects that
can be seen to great distances. Presently, measurements of supernova type Ia
up to $z\sim 1.8$ are available. The fact that these very distant supernova
appear dimmer than would be expected in a pure matter-dominated universe led
to the conclusion that the universe is accelerating. Much speculation has
arisen as to the source of this acceleration and even its reality has been
questioned. But, the fact that the spectrum of anisotropy of the cosmic
microwave background radiation indicates that the spatial section of the
geometry of the universe must be almost flat, while the observations of the
virialized system (galaxies, clusters of galaxies, etc.) indicate a low
density of the universe, implies indirectly the acceleration of the
universe. In fact, in order to complete the cosmic energy budget and have
flat spatial sections, it is necessary to include a component that does not
agglomerate locally, but remains as a smooth component of universe. To have
this feature, this component must have negative pressure, and consequently
it must dominate the matter content of the universe asymptotically, driving
the accelerated expansion in the later phases of the cosmic evolution.

In brief, to explain the acceleration of the universe, an exotic component
in the cosmic budget exhibiting negative pressure is needed. This exotic
component is named \textit{dark energy}. We must remember that observations
require also a second non-baryonic component, called \textit{dark matter},
with zero pressure, necessary to explain conveniently the formation of
structures in the universe and the dynamics of local, virialized system,
like galaxies and clusters of galaxies. However, while there exists a lot of
reasonable candidates to represent dark matter (neutralinos, axions, sterile
neutrinos, primordial black holes etc -- for a review, see reference~\cite{BHS}),
it is not clear what kind of fluid or field would constitute dark energy.
The first natural candidate to be evoked has been the cosmological constant,
seen as a phenomenological manifestation of the vacuum energy of quantum
fields existing in the universe. However, this very attractive possibility
is plagued with fine -tunings and poorly understood issues, (see the
classical references~\cite{BT}). In spite of this, the cosmological constant
remains the most popular candidate to represent dark energy, leading to the
so-called $\Lambda $CDM model, highly supported by observations. A simple
extension of general relativity, created by restricting the variational
principle for deriving the Einstein equations to causal variations, leads to
a general prediction that there is a cosmological constant with the observed
value and a prediction that the value of the dimensionless curvature will be
observed to be -0.0055~\cite{BS}.

The cosmological constant implies an equation of state such that $\alpha
_{x}=p_{x}/\rho _{x}=-1$, where $\rho _{x}$ is the density and $p_{x}$ is
the pressure (the subscript $x$ denotes a dark energy component).
Observations will never give an exactly precise result for the equation of
state of dark energy (or any component of the universe), due to statistical
and systematic errors. But, if these dispersions in the evaluation of the
dark energy equation of state can be reduced substantially, a clear case for
the simple cosmological constant can emerge. We are far from this situation.
But, in the search for the determination of $\alpha _{x}$ some curious
results have appeared: while the estimations lead to values near $\alpha
_{x}=-1$, more negative values are highly admitted. This fact has raised a
dramatic speculation: could dark energy have an equation of state such that $%
\alpha _{x}<-1$? In fact, for example, if $\alpha _{x}$ is constant and the
spatial sections of the universe is flat, the recent results of 7-years WMAP
observations indicate that $\alpha _{x}=-1.10\pm 0.14$ ~\cite{K} at $1\sigma $.

If $\alpha _{x}<-1$, all energy conditions are violated, in particular the
null energy condition, which requires $\rho _{x}+p_{x}\geq 0$. From the
conservation of the energy-momentum tensor in an expanding universe, we
obtain the equation, 
\begin{equation}
\dot{\rho}_{x}+3\frac{\dot{a}}{a}(\rho _{x}+p_{x})=0,
\end{equation}%
a where $a$ is the scale factor of the universe, and dots mean derivatives
with respect to the cosmic time. From this expression, it comes out that the
violation of the null energy condition implies that the energy density of
the dark energy fluid grows with the expansion of the universe, instead of
decreasing. A remarkable consequence of such behaviour is that any universe
dominated by such an exotic fluid will inevitably hit a singularity in its
future evolution, after a finite proper time. At this future singularity,
the expansion scale factor $a$ and the energy density $\rho _{x}$ both
diverge.  This highly singular state in the future evolution of the universe
has been called \textit{big rip singularity}~\cite{C}, and it is a remarkable and
plausible example of what are called \textit{future cosmological
singularities}.

However, the situation is more complicated, and future singularities can
exist even if the energy conditions are not violated. The possibility of
future singularities was pointed out for the first time in reference~\cite{BGT} and
their occurrence does not necessarily require the violation of the energy
conditions. There is also a later discussion in reference~\cite{SSS}. A recent example of
future singularities which does not violate the null energy condition is the
'big brake' singularity, which emerges from the DBI action~\cite{GKMP,KGGMK}. The big
brake singularity has the curious property that it can be traversed by a
pointlike particle~\cite{KGLGP}. A milder type of finite time singularity is the
so-called "\textit{sudden} singularity" \cite{B}. The \textit{sudden} future
singularity occurs without violating any energy condition (so $\rho +p\geq 0$
and $\rho +3p\geq 0$ at all times). This singularity is characterized by a
finite value for the scale factor, its first time derivative, and the
density, while the second derivative of the scale factor and the pressure
diverge at finite time~\cite{B}. They are singularities of the weak sort
discussed by Tipler~\cite{T} and Krolak~\cite{Kr}. They cannot occur if $dp/d\rho $
is continuous.

In general, it is believed that the fate of the universe near any
singularity (past or future) must be affected by quantum effects arising in
the extreme conditions that exist in its spacetime neighborhood. In the case
of the \textit{big rip}, this problem has been treated for example, in
references~\cite{BFH,ABFH,BA}. In these investigations, it was found that the quantum
effects are important. But, the conclusions concerning the back reaction of
the quantum effects on the evolution of the universe were harder to decide
unambiguously. In the case of the \textit{sudden} singularity, quantum
effects were studied in references~\cite{BBFH,BBDFH,H}, and the results indicated that
quantum effects do not change the evolution near the singularity. These
results were obtained for massive and massless scalar fields.

Here, the fate of quantum effects near future singularities is reviewed. We
begin, in next section, by describing the some general class of future
singularities. In section~3 the evolution of a scalar field in the
background of \textit{big rip} and \textit{sudden} future singularities is
determined. In section 4, the structure of the background solution is shown,
and the solution for the scalar field equation is found. In section 5,
quantum effects near future singularities are studied. In section 6 we
present our conclusions.

\section{Future singularities}

Generally, a singularity in geometric theories of gravitation can be
characterized by the divergence of some curvature invariants or,
alternatively, by the incompleteness of the geodesic trajectories. Cosmology
is commonly studied using a homogeneous and isotropic space-time. In this
case, all dynamics is encoded in the behavior of a single function, the
scale factor $a(t)$, determined by some matter density $\rho (t)$ endowed
with a pressure $p(t)$. In this case, singularities appear as a divergence
in the Hubble factor $H=\frac{\dot{a}}{a}$ and in the density $\rho (t)$. A
traditional example is the initial \textit{big bang} singularity, for which 
$H(t)\rightarrow \infty $ and $\rho (t)\rightarrow \infty $ as 
$t\rightarrow 0$. Of course, at the big bang singularity the curvature scalars diverge and
the geodesics "begin" in the singularity, implying the the space-time is
geodesically incomplete.

The future singularities which occur at the end of a cosmological evolution,
have in principle many of the features as the \textit{big bang} singularity,
including the divergence of the curvature invariants, which are displayed by
many types of singularity. However, geodesic incompleteness is not a
universal characteristic of these singularities and nor is a divergence in
the Hubble parameter or in the density. The \textit{big rip} singularity,
which requires the violation of all energy conditions, bears a close
resemblance to the \textit{big bang} singularity, since, besides the
divergence of curvature invariants, it is incomplete geodesically, and the  
energy density and the Hubble parameter both diverge. This is direct
consequence of the violation of the null energy condition. As already
remarked in the introduction, since $\rho +p$ is negative when the null
energy condition is violated, the energy density \textit{grows} as the
universe expands, in contrast to the usual situation. As consequence, $%
R\rightarrow \infty $, $H\rightarrow \infty $ and $\rho \rightarrow \infty $
after a finite time $t_{s}$. Moreover, the geodesics terminate at this
singularity, and are inextendible.

The case of the \textit{big rip} singularity can be described in a simple
way by assuming a fluid (named from now on \textit{phantom fluid}) with an
equation of state of the type 
\begin{equation}
p=\alpha \rho ,\quad \alpha <-1.
\end{equation}%
Solving the conservation equation, the energy density scales as 
\begin{equation}
\rho \propto a^{-3(1+\alpha )}.
\end{equation}%
Since, $\alpha <-1$, $\rho $ grows as $a$ increases.

This singularity occurs in a finite proper time. This can be seen more
simply in the following way. In terms of the cosmic time, the scale factor
behaves as 
\begin{equation}
a \propto t^\frac{2}{3(1 + \alpha)}.
\end{equation}
Since $1 + \alpha$ is negative, this solution can represent an expanding
universe if $- \infty < t < 0_-$. It must be remarked that at $t \rightarrow
- \infty$ there is no singularity, as we will see below. If we live in a
moment $t_0 < 0$, the time to elapse from $t = t_0$ to $t = 0_-$ is
obviously finite. If there is any other form of "normal" matter, the phantom
fluid will always dominate asymptotically, since the density of normal forms
of matter decreases with the expansion. In reference~\cite{K} a simple model
including pressureless matter and phantom fluid has been described. On the
other hand, since $R \propto \frac{1}{t^2}$, $\rho \propto \frac{1}{t^2}$
and $H \propto \frac{1}{t}$, all these quantities diverge as $t \rightarrow
0_-$. It is in this sense that the \textit{big rip} singularity can be
considered as the reverse of the \textit{big bang} singularity, with a
Minkowskian asymptotic space-time as $a \rightarrow 0$ and a singular state
as $a \rightarrow \infty$.

There are other classes of mild future singularities. We understand a "mild
singularity" to be a singularity that exhibits a divergence in the curvature
invariants (which is a requirement), but perhaps with no divergence in the
density. If the density does not diverge then Einstein's equations imply
that the Hubble function also does not diverge at the singularity. At same
time, since density is connected with the Hubble function, it must reach a
finite value at singularity. However, the Ricci scalar is given by 
\begin{equation}
R=-6\biggr[\frac{\ddot{a}}{a}+\biggr(\frac{\dot{a}}{a}\biggl)%
^{2}\biggl].
\end{equation}%
In order to have a divergence in $R$, we see that $\ddot{a}$ must diverge.
Through Einstein's equations, this implies that the pressure must diverge.
This curious structure may allow that the geodesics to be continued through
the singularity.

One example of these class of \textquotedblleft mild future
singularities\textquotedblright\ is the \textit{sudden }singularity, which
can be described by the following expression for the scale factor~\cite{B,B2,BT2}: 
\begin{equation}
a(t)=\biggr(\frac{t}{t_{s}}\biggl)^{q}(a_{s}-1)+1-\biggr(1-\frac{t}{t_{s}}%
\biggl)^{n},  \label{sf-sud}
\end{equation}%
where $t_{s}$ is the time where the \textit{sudden} singularity occurs, and $%
a_{s}$ is the value of the scale factor at this moment. Moreover, $0<q\leq 1$
and $1<n<2$ where $q$ and $n$ are free constants and no specific relation is
assumed between the pressure $p$ and the density $\rho $.

The \textit{sudden} singularity does not require any violation of the energy
conditions. Pressure can remain positive. The only requirement is the
divergence of pressure in a given finite time. One difficult with such
scenario is to find a type of matter that could satisfy these requirements.
While this seems to be difficult using ordinary types of fluids,
non-standard self-interacting scalar fields may fulfill the requirements
needed for a \textit{sudden} singularity scenario.

In fact, a self-interacting scalar field which presents non-standard
coupling structures may lead to new and unexpected scenarios. One example is
given by the DBI action, 
\begin{equation}
\mathcal{L}=\sqrt{-g}V(T)\sqrt{1-T_{;\rho }T^{;\rho }},
\end{equation}%
where $T$ is a \textit{tachyonic} scalar field, and $V(T)$ is a potential
term. This action can emerge from some specific configuration of string
theory. The choice of the potential $V(T)$ may lead to very new kind of
cosmological scenario. One example has been given in~\cite{SSS}, where a new kind
of future singularity has been exhibited: the \textit{big brake singularity}%
. In this singularity, the Hubble parameter vanishes on the singularity,
while the second derivative of the scale factor goes to $-\infty $, from
which the name \textit{big brake}. Geodesics can also traverse the \textit{%
big brake} singularity.

From now on, we will concentrate our analysis on the \textit{big rip} and 
\textit{sudden} singularities.

\section{The master equation}

We will consider the quantum creation of particles near a given future
cosmological singularity. Two types of future singularity will be analyzed:
the \textit{big rip} singularity and the \textit{sudden} singularity. Let us
initially consider a general massive, non-minimally coupled scalar field $%
\phi $, giving by the following Lagrangian: 
\begin{equation}
\mathcal{L}=\frac{1}{2}\phi _{;\rho }\phi ^{;\rho }-\frac{1}{2}m^{2}\phi
^{2}+\frac{\xi }{2}R\phi ^{2}.  \label{lagran}
\end{equation}%
The conformal coupling corresponds to $\xi =\frac{1}{6}$. From the
Lagrangian (\ref{lagran}) we can deduce the following field equation: 
\begin{equation}
\Box \phi +m^{2}\phi -\xi R\phi =0.
\end{equation}%
For a flat FLRW metric, this equation reduces to 
\begin{equation}
\phi ^{\prime \prime }+2\frac{a^{\prime }}{a}\phi ^{\prime }+\biggr\{%
k^{2}+m^{2}a^{2}+6\xi \frac{a^{\prime \prime }}{a}\biggl\}\phi =0.
\label{kga}
\end{equation}

In general, at future singularities the second derivative of the scale
factor diverges, while the first one remains finite. Hence, it is convenient
to transform equation (\ref{kga}) in order to eliminate this possibly
singular term. This can be achieved by defining $\phi =a^{-6\xi }\chi $.
This transformation leads to the equation 
\begin{eqnarray}
&\chi ^{\prime \prime }&+2\biggr(1-6\xi \biggl)\frac{a^{\prime }}{a}\chi
^{\prime }  \notag  \label{kgb} \\
&+&\biggr\{k^{2}+m^{2}a^{2}+6\xi (6\xi -1)\biggr(\frac{a^{\prime }}{a}\biggl)%
^{2}\biggl\}\chi =0.
\end{eqnarray}%
The equation (\ref{kgb}) simplifies considerably for a conformal coupling, $%
\xi =\frac{1}{6}$. If, a massless field is also considered, equation (\ref%
{kgb}) takes the form of an harmonic oscillator equation.

Particle production near the \textit{big rip} has been analyzed the
references~\cite{Kr,BFH,ABFH}, considering only massless scalar particle, while the
corresponding analysis for the \textit{sudden} singularity has been
performed in reference~\cite{BA}. In reference~\cite{BBFH} the massive case with
conformal coupling in the background of the \textit{sudden} singularity has
been studied. We will review the results obtained in these references later.
We must remark that the non-minimal coupling should strengthen any quantum
effects near the singularity, and also introduces some technical features
that leads to exact solutions for the problem.

Under variations with respect to the metric, the Lagrangian (\ref{lagran})
gives the following momentum-energy tensor: 
\begin{eqnarray}
T_{\mu \nu } &=&(1-2\xi )\phi _{;\mu }\phi _{;\nu }+\biggr(2\xi -\frac{1}{2}%
\biggl)g_{\mu \nu }\phi _{;\rho }\phi ^{;\rho }  \notag \\
&+&\frac{1}{2}m^{2}g_{\mu \nu }\phi ^{2}+\xi G_{\mu \nu }\phi ^{2}  \notag \\
&-&2\xi \phi \biggr(\phi _{;\mu ;\nu }-g_{\mu \nu }\Box \phi \biggl),
\end{eqnarray}%
where 
\begin{equation}
G_{\mu \nu }=R_{\mu \nu }-\frac{1}{2}g_{\mu \nu }R
\end{equation}%
is the Einstein tensor. For a conformal coupling ($\xi =\frac{1}{6}$), this
expression reduces to, 
\begin{eqnarray}
T_{\mu \nu } &=&\frac{2}{3}\phi _{;\mu }\phi _{;\nu }-\frac{1}{6}g_{\mu \nu
}\phi _{;\rho }\phi ^{;\rho }+\frac{1}{2}m^{2}g_{\mu \nu }\phi ^{2}  \notag
\label{energia-massa-chi} \\
&&-\frac{\phi }{3}\biggr(\phi _{;\mu ;\nu }-g_{\mu \nu }\Box \phi \biggl)+%
\frac{1}{6}G_{\mu \nu }\phi ^{2}.
\end{eqnarray}%
In the case of minimally coupled massive field ($\xi =0$), the
energy-momentum tensor reduces to 
\begin{equation}
T_{\mu \nu }=\phi _{;\mu }\phi _{;\nu }-\frac{1}{2}g_{\mu \nu }\phi _{;\rho
}\phi ^{;\rho }+\frac{1}{2}m^{2}g_{\mu \nu }\phi ^{2}. \label{energia-massa-schi}
\end{equation}

From now on, we will consider two main cases: $m\neq 0$ and $\xi =\frac{1}{6}
$ for the massive conformally coupled scalar field, and $m=0$ and $\xi =0$,
for the massless minimally coupled scalar field.

\section{The cosmological background and the solutions of the master equation%
}

The first task in analyzing the quantization of the scalar field equations
is to determine the background where the quantum field evolves. As stated
above, two cases will be considered: the \textit{big rip} and the \textit{%
sudden} singularity.

\subsection{The \textit{big rip} background}

Let us consider the flat Friedmann equation and the conservation law for a
fluid with an equation of state $p=\alpha \rho $: 
\begin{eqnarray}
\biggr(\frac{a^{\prime }}{a}\biggl)^{2} &=&\frac{8\pi G}{3}\rho a^{2}, \\
\rho ^{\prime }+3(1+\alpha )\rho  &=&0.
\end{eqnarray}%
The general solution for the scale factor is given by $a=a_{0}|\eta |^{\frac{%
2}{1+3\alpha }}$ ($\eta $ is the conformal time defined by $dt=a(\eta )d\eta 
$), while the density behaves as $\rho =\rho _{0}a^{-3(1+\alpha )}$. If $%
\alpha <-\frac{1}{3}$ then accelerated expansion occurs. Moreover, if $%
\alpha <-1$ (phantom regime) then the matter density grows during this
accelerated expansion, leading to a \textit{big rip}. One of the main
aspects of these accelerating solutions is that the conformal time takes
values in the interval $-\infty >\eta >0_{-}$. Note that for $\eta
\rightarrow -\infty $ the scale factor goes to zero, and the Minkowski
space-time is asymptotically approached in the phantom regime, since $\rho
\rightarrow 0$ in this regime. This property will be important in order to
fix the initial conditions.

\subsection{\textit{Sudden} singularity: two cosmological eras}

For the \textit{sudden} singularity, the situation is more complex due to
the complicated form of the scale factor. But, the scale factor (\ref{sf-sud}%
) admits two asymptotic forms, which we will call the \textit{primordial
phase} and the \textit{singular phase, }respectively.

\begin{itemize}
\item Primordial phase, $t \rightarrow 0$: 
\begin{eqnarray}
a &\rightarrow& \biggr(\frac{t}{t_s}\biggl)^q(a_s - 1), \\
\dot a &\rightarrow& \frac{q}{t_s}\biggr(\frac{t}{t_s}\biggl)^{q-1}(a_s - 1),
\\
\ddot a &\rightarrow& \frac{q}{t_s^2}(q - 1)\biggr(\frac{t}{t_s}\biggl)%
^{q-2}(a_s - 1).
\end{eqnarray}

\item Singular phase, $t \rightarrow t_s$: 
\begin{eqnarray}
a &\rightarrow& a_s, \\
\dot a &\rightarrow& \frac{q}{t_s}(a_s - 1), \\
\ddot a &\rightarrow& -\frac{n}{t_s^2}(n - 1)\biggr(1 - \frac{t}{t_s}\biggl)%
^{n-2}.
\end{eqnarray}
\end{itemize}

There will be a radiation-dominated primordial phase if $q=1/2$. On the
other hand, in the singular phase, the scale factor and its first derivative
approach constant values, and the second derivative, $\ddot{a}$, diverges as 
$t\rightarrow t_{s}$, since $n<2$.

In terms of the conformal time, $d\eta =a^{-1}dt\,$, we have for the scale
factor evolution to leading order:

\begin{itemize}
\item Radiation-dominated primordial phase: 
\begin{equation}
a=a_{0}\eta .
\end{equation}

\item Singular phase: 
\begin{equation}
a = a_s.
\end{equation}
\end{itemize}

The scale factor and its first derivative must be continuous during the
transition from one phase to the other. If $a_{s}$ is the scale factor value
at the moment of the transition, and $H_{0}$ the corresponding Hubble
parameter value, then the transition moment is given by $\eta
_{t}=1/(H_{0}\,a_{s})$ and $a_{0}=H_{0}\,a_{s}^{2}$.

The isotropic and homogeneous form we have assumed for the cosmological
evolution of the scale factor, $a(t)$, towards a \textit{sudden} singularity
captures the essential features of the general solution of the Einstein
equations near such a singularity. For some more general scenarios,
including anisotropies, see refs~\cite{BT2,BCT,BL,S,CST,CK}.

We are ready now to determine the solution for the master equations for some
relevant configurations.

\subsection{\textit{Big rip}: massless, minimally coupled case}

When $\xi =m=0$ it is more practical to solve directly the equation for the
field $\phi $. The Klein-Gordon equation assumes the form, 
\begin{equation}
\phi ^{\prime \prime }+2\frac{a^{\prime }}{a}\phi ^{\prime }+k^{2}\phi =0,
\label{kg-1}
\end{equation}%
where from now on $k$ denotes the wave number related to the Fourier
decomposition of the spatial dependence of the scalar field. When the scale
factor is given by $a=a_{0}|\eta |^{\frac{2}{1+3\alpha }}$, representing the 
\textit{big rip} scenario, the solution is: 
\begin{equation}
\phi _{k}(\eta ,\vec{x})=c_{1}\eta ^{\nu }H_{\nu }^{(1)}(k\eta )e^{i\vec{k}%
\cdot \vec{x}},  \label{rip-sol}
\end{equation}%
where $c_{1}$ is an integration constant, independent of $k$, $\nu =-\frac{%
3(1-\alpha )}{2(1+3\alpha )}$ and $H_{\nu }^{(1)}(x)$ is the Hankel function
of the first kind of order $\nu $. This form of the final solution, which
should contain in principle two independent functions and coefficients with
dependence on $k$, is dictated by the imposition of an initial Bunch-Davies
vacuum state~\cite{BD}.

\subsection{\textit{Sudden} singularity: massless, minimally coupled case}

Let us return to the Klein-Gordon equation (\ref{kg-1}). For the \textit{%
sudden} singularity, we have previously considered  two phases, a radiative
phase and a singular phase. The initial conditions, corresponding to the
Bunch-Davies vacuum state, are imposed at the beginning of the radiative
phase, where the solution of the Klein-Gordon equation may be expresses in
the form of plane waves: 
\begin{equation}
\phi _{k}(\eta )=\frac{e^{ik\eta }}{\sqrt{2k}}.
\end{equation}%
But, the main interested is in the behavior of the quantum field near the
singularity. In this phase, the solution of equation (\ref{kg-1}) reads, 
\begin{equation}
\phi (\eta ,\vec{x})=e^{-\eta }\biggr\{A_{+}e^{i(\omega \eta -\vec{k}\cdot 
\vec{x})}+A_{-}e^{-i(\omega \eta +\vec{k}\cdot \vec{x})}\biggl\},
\end{equation}%
where $\omega =\sqrt{k^{2}-1}$ and the constants $A_{\pm }$ are fixed by the
matching conditions across the two phases: 
\begin{equation}
A_{\pm }=\frac{1}{2\omega }\sqrt{\frac{1}{2k}}e^{i(k\mp \omega )+1}(\omega
\pm k).
\end{equation}

\subsection{\textit{Sudden} singularity: massive, conformally coupled case}

If the mass is non-zero, and $\xi =\frac{1}{6}$, the resulting Klein-Gordon
equation takes the following form during the singular phase: 
\begin{equation}
\chi ^{\prime \prime }+(k^{2}+m^{2}a^{2})\chi =0.
\end{equation}%
where $\chi =a\phi $. During the radiative phase it is reasonable to
consider the massless approximation, since we are in the regime where $%
a\rightarrow 0$. During the singular phase, the equation is 
\begin{equation}
\chi ^{\prime \prime }+(k^{2}+m^{2}a_{0}^{2}\eta ^{2})\chi =0.  \label{eq}
\end{equation}%
which again has plane wave solutions with a modified frequency. Hence, the
solutions during the two phases can be written as follows: 
\begin{eqnarray}
\phi _{k}(\eta ) &=&\frac{e^{ik\eta }}{\sqrt{2k}}\quad \quad %
\mbox{(primordial phase)},  \label{s1m} \\
\phi _{k}(\eta ) &=&\xi _{01}e^{i\tilde{\omega}\eta }+\xi _{02}e^{-i\tilde{%
\omega}\eta }\quad \quad \mbox{(singular phase)}, \label{s2m}
\end{eqnarray}%
where $\xi _{01,02}$ are constants, to be fixed by the matching conditions,
and $\tilde{\omega}=\sqrt{k^{2}+m^{2}a_{0}^{2}}$.

\section{The energy of the created particles and the regularization procedure%
}

The $0-0$ component of the energy-momentum tensor gives the energy density
of a given configuration. The solutions shown in the previous section give
the expression for the scalar field (massive or massless) for a given mode $k
$. Inserting this expression in the energy-momentum tensor, the
corresponding $0-0$ component gives the energy associated with this mode.
The total energy is obtained by integrating over all modes. In general, this
integration leads to a divergent quantity, a common problem in quantum field
theory. Hence, in order to make sense of the energy associated with the
field, a renormalisation procedure must be employed. The renormalisation can
be interpreted as a redefinition of the fundamental constants present in the
original problem. In order for this procedure to be physically meaningful,
it is essential that the final result does not depend on how the
renormalisation is performed.

Hence, our problem here is to compute the expression 
\begin{equation}
\rho =\int \rho _{k}d^{3}k,
\end{equation}%
with $\rho _{k}={T}_{0}^{0}$. More precisely, we are interested in the final
expression for the renormalized energy, $\rho ^{ren}$. It is this expression
that will give the energy associated with any particles created by the
quantum processes near the singularity, and which may be relevant to the
computation of any associated back-reaction phenomena in the evolution of
the universe.

In order to regularize the expression of the energy, we use the $n$-wave
method exposed in the reference~\cite{ZS}. This method is based on the
Pauli-Villars technique used for quantum field theory in Minkowski
space-time. First, let us write the energy as, 
\begin{equation}
\rho =\int_{0}^{\infty }\rho _{k}(k,m)k^{2}\,dk.
\end{equation}%
Let us define, 
\begin{equation}
^{{}}\rho _{k}^{(n)}\equiv \frac{1}{n}\rho _{k}(nk,nm),
\end{equation}%
where $n$ is a parameter that characterizes the order of the divergence.
From this expression we construct the quantities, 
\begin{equation}
E_{k}^{p}=\lim_{n\rightarrow \infty }\frac{\partial ^{p}\rho _{k}^{(n)}}{%
\partial (n^{-2})^{p}}.
\end{equation}%
The expression for the regularised energy is given by, 
\begin{equation}
\rho _{k}^{reg}=\rho_{k}-E_{k}^{0}-E_{k}^{1}-\frac{1}{2}E_{k}^{2},
\end{equation}%
where $E_{k}^{0}$ eliminates the logarithmic divergence, $E_{k}^{1}$ the
quadratic divergence, and $E_{k}^{2}$ the quartic divergence -- all those
that are normally present in the energy-momentum tensor. This regularization
of the energy corresponds to a full renormalisation of the coupling
constants, as described in~\cite{GMM,BLM}.

The goal now is to determine the final expression for the regularized
energy-momentum tensor, especially for the energy density that corresponds
to the component $0-0$ of this tensor.

\subsection{\textit{Big rip}: the massless, minimally coupled case}

Using the solution (\ref{rip-sol}) and the expression for the energy given
by the $0-0$ component of the energy momentum tensor (\ref%
{energia-massa-schi}) with $m = 0$, we obtain, 

\begin{eqnarray}
\rho _{k} &=&A\eta ^{\gamma -3}x^{2}\Big[H_{\nu -1}^{(1)}(x)\,H_{\nu
-1}^{(2)}(x)  \notag \\
&+&H_{\nu }^{(1)}(x)\,H_{\nu }^{(2)}(x)\Big].
\end{eqnarray}%
The integration on all $k$ modes reveals the existence of logarithmic,
quadratic and quartic divergences. Using the $n$-wave regularization scheme,
we write 
\begin{eqnarray}
\rho ^{ren} &=&\int_{0}^{\infty }x^{2}(\rho _{k}-E_{k}^{0}-E_{k}^{1})dx 
\notag  \label{rip-ren} \\
&-&\int_{1/\sigma }^{\infty }x^{2}E_{k}^{log}dx.
\end{eqnarray}%
The last term is a modification of usual method in order to include the
logarithmic divergence in the massless case, see~\cite{Kr}. The result is: 
\begin{equation*}
\rho ^{ren}=\bar{A}_{1}\eta ^{\gamma -3}I_{1}=\bar{A}_{1}\eta ^{\frac{%
-12(1+\alpha )}{1+3\alpha }}I_{1},
\end{equation*}%
where $I_{1}$ is a number. This expression describes how the energy of the
created particles evolves with time. 

Taking the ratio between created particles and the background phantom fluid $%
\rho _{x}$, we find 
\begin{equation}
\frac{\rho ^{ren}}{\rho _{x}}\propto \eta ^{-6\left( \frac{1+\alpha }{%
1+3\alpha }\right) },
\end{equation}%
which diverges as the singularity at $\eta =0_{-}$ is approached. This may
indicate that the evolution of the universe is modified by quantum effects
and the singularity is avoided. But, in order to verify such effect, a full
back-reaction analysis is necessary. Similar studies for the \textit{big rip}
singularity have been made in the context of quantum cosmologies, with
inconclusive results, see references~\cite{DKS,KKS,BL2,PNP}.

\subsection{\textit{Sudden} singularity for the massless, minimally coupled
case}

Let us turn now to the \textit{sudden} singularity case. As in the previous
analysis for the \textit{big rip}, we consider a massless, minimally coupled
scalar field. In contrast to the \textit{big rip} analysis above, an
approximation considering the two phases is now necessary, and the initial
vacuum condition is imposed in the first phase, which is that of a
radiation-dominated universe. Using the solutions for the Klein-Gordon for
the two phase given above, we find that the total energy is given by the
following integral: 
\begin{eqnarray}
\rho =Ae^{y}\int_{0}^{\infty }dk\frac{k}{\omega ^{2}} &\biggr\{%
&(2k^{2}-1)k^{2}  \notag \\
&-&\cos \omega y+\omega \sin \omega y\biggl\},
\end{eqnarray}%
where $y=2(1-\eta )$ and $\omega =\sqrt{k^{2}-1}$. The background constants
are fixed such the singularity occurs at $\eta =1$, $y=0$.

Employing the $n$-wave regularization scheme, the regularized energy can be
determined~\cite{ABFH}: 
\begin{equation}
\rho ^{ren}=\bar{A}e^{y}\biggr\{Chi(-y)+\frac{\cosh y}{y}\biggl\},
\end{equation}%
where $Chi$ denotes the hyperbolic cosine integral function. The regularized
energy decreases as the singularity is approached ($y\rightarrow 0$). Hence,
the quantum effects are ineffective in preventing the singularity, at least
for the massless scalar field case.

\subsection{\textit{Sudden} singularity for the massive, conformal coupling}

In this case, as for the previous one, two phases were considered. The
initial vacuum state is fixed during the first phase, and the energy is
computed during the second phase, in order to evaluate the possibility of a
back-reaction effect on the evolution of the universe.

Using the solutions (\ref{s1m},\ref{s2m}), the matching conditions and the
expression for the energy-momentum tensor (\ref{energia-massa-chi}) we
obtain for the energy of the $k$ mode: 
\begin{equation}
\rho _{k}=\frac{k}{4}\biggr(1-\frac{k}{\tilde{\omega}}\biggl).
\end{equation}%
An integration over all $k$-modes gives, 
\begin{equation}
\rho =\int_{0}^{\infty }\rho _{k}\,d^{3}k=\pi \int_{0}^{\infty }k^{2}\tilde{%
\omega}\biggr(1-\frac{k}{\tilde{\omega}}\biggl)^{2}dk.  \label{rhoef}
\end{equation}%
This expression clearly diverges so it is necessary to regularise it. But,
heuristically, since it is a polynomial expression, it seems clear that
after regularisation we must obtain zero. Hence, the particle production
should not contribute to the energy-momentum tensor and the \textit{sudden}
singularity is unaffected by these quantum effects

Note that the integral (\ref{rhoef}) admits an analytical solution: 
\begin{eqnarray}
\int\rho_k\,d^3k &=& \pi \int k^2\tilde\omega\biggr(1 - \frac{k}{\tilde\omega%
}\biggl)^2dk  \notag \\
&=& \pi\biggr\{k\sqrt{k^2 + \bar m^2}\biggr(\frac{k^2}{2} - \frac{\bar m^2}{4%
}\biggl) - \frac{k^4}{2}  \notag \\
&+& \frac{\bar m^2}{4}\ln\biggr[2\biggr(k + \sqrt{k^2 + \bar m}\biggl)\biggr]%
\biggl\},
\end{eqnarray}
with $\bar m = m\,a_0$. There is no infrared divergence, but there is a
logarithmic divergence when $k \rightarrow \infty$ (ultraviolet limit).

We have, 
\begin{equation}
\rho _{k}=\sqrt{k^{2}+\bar{m}^{2}}-2k+\frac{k^{2}}{\sqrt{k^{2}+\bar{m}^{2}}}.
\end{equation}%
It follows that 
\begin{equation}
\rho _{k}^{(n)}=\rho _{k}.
\end{equation}%
Hence, only the zero-order term survives, and leads to, 
\begin{equation}
\rho _{k}^{ren}=\rho _{k}-E_{k}^{0}=\rho _{k}-\rho _{k}=0.
\end{equation}%
As we suspected, the renormalized energy is zero. There is no effect, and
the quantum phenomena associated with the cosmological dynamics do not
change the character of the \textit{sudden} singularity or prevent its
occurrence.

\section{Conclusions}

It is known that quantum effects may play an important r\^{o}le near
classical singularities, for example, for the \textit{big bang} cosmological
scenario. Even if the back reaction of these quantum effects on the
classical evolution is still an open question, there are at least hints that
quantum effects may lead to a dramatic deviation from the classical behavior~\cite{BD}.

Recently, new kinds of singularities have been identified in cosmology, the
so-called future singularities. While the \textit{big bang} singularity
occurs in the origin of the universe, future singularities may mark the end
of the universe. The interest in this kind of singularity has increased
recently because of the unusual possibility that the presently observed
accelerated expansion of the universe may be driven by a phantom fluid,
which violates all energy conditions. However, it had been shown some time
earlier that such future singularities may occur even if the energy
conditions are not violated.

The fate of quantum effects near future singularities has been reviewed
here, with special attention to the \textit{big rip} singularity and the 
\textit{sudden} singularity. Concerning the \textit{big rip,} the quantum
effects may be relevant, but a more careful analysis of the back reaction
process is necessary to decide under what conditions the \textit{big rip}
can be evaded by quantum effects~\cite{BA}. In the case of the \textit{sudden}
singularity, the results obtained so far indicate that this singularity is
robust against quantum effects because they are negligible in its vicinity.

It must be remarked however that such studies has been carried out using
special configurations of scalar fields. More general quantum fields must be
analyzed and, of course, the problem of the back reaction must also be
treated in greater generality.

\vspace{0.5cm} \noindent \textbf{Acknowledgements:} We thank CNPq (Brasil)
for partial financial support. JDB thanks S. Cotsakis, S.Z.W. Lip, C.G.
Tsagas and A. Tsokaros for their active collaboration.

\bibliographystyle{plain}

\end{document}